\documentclass[pre,twocolumn]{revtex4} 
\usepackage{amsmath,epsfig,amssymb}

\newcommand{\be}{\begin{equation}}
\newcommand{\ee}{\end{equation}}
\newcommand{\bp}{\begin{picture}}
\newcommand{\ep}{\end{picture}}
\newcommand{\ba}[1]{\begin{array}{#1}}
\newcommand{\ea}{\end{array}}
\newcommand{\bea}{\begin{eqnarray}}
\newcommand{\eea}{\end{eqnarray}}
\newcommand{\ho}{\hat{\omega}}

\newcommand{\bfm}[1]{\mbox{\boldmath $#1$}}
\newcommand{\tr}{\mbox{tr}}

\begin{document}

\title{Entropy-induced Microphase Separation in Hard Diblock Copolymers}
\author{Paul P.\ F.\ Wessels}
\email{wessels@thphy.uni-duesseldorf.nl}
\altaffiliation[Present address:]{
Heinrich-Heine-Universit\"{a}t
D\"{u}sseldorf,
Institut f\"{u}r Theoretische Physik II
Universit\"{a}tsstra{\ss}e 1, Geb\"{a}ude 25.32,
D-40225 D\"{u}sseldorf, Germany}
\author{Bela M.\ Mulder}
\affiliation{FOM Institute for Atomic and Molecular Physics,
Kruislaan 407, 1098 SJ Amsterdam, The Netherlands}
\date{\today}

\begin{abstract}
Whereas entropy can induce phase behavior that is as rich as seen in
energetic systems, microphase separation remains a very rare phenomenon in entropic systems.
In this paper, we present a density functional approach to study the possibility
of entropy-driven microphase separation in diblock copolymers. Our
model system consists of copolymers composed of freely-jointed
slender hard rods. The two types of monomeric segments have
comparable lengths, but a significantly different diameter, the
latter difference providing the driving force for the phase
separation. At the same time these systems can also exhibit liquid
crystalline phases. We treat this system in the appropriate
generalization of the Onsager approximation to chain-like
particles. Using a linear stability (bifurcation) analysis, we
analytically determine the onset of the microseparated and the
nematic phases for long chains. We find that for very long chains
the microseparated phase always preempts the nematic. In the limit
of infinitely long chains, the correlations within the chain
become Gaussian and the approach becomes exact. This allows us to
define a Gaussian limit in which the theory strongly simplifies
and the competition between microphase separation and liquid
crystal formation can be studied essentially analytically. Our
main results are phase diagrams as a function of the effective
diameter difference, the segment composition and the length ratio
of the segments. We also determine the amplitude of the positional
order as a function of position along the chain at the onset of
the microphase separation instability.
Finally, we give suggestions as to how this type of entropy-induced
microphase separation could be observed experimentally.
\end{abstract}

\maketitle

\section{Introduction}
\label{sec:intro}

Microphase separation (MPS) is the phenomenon where an initially
homogeneous phase develops an inhomogeneous spatial structure on a
microscopic scale. Usually such systems consist in part of
thermodynamically incompatible components that left by themselves
would tend to (macroscopically) phase separate. However, due to
additional constraints of a physical or chemical nature the
spatial separation between the components is prevented from
increasing beyond a microscopic length scale. This leads to phases
in which the components can demix only locally. There are a few
archetypical examples of systems showing MPS: (i) Two (usually
flexible) polymers species that have an unfavourable mutual
interaction energy which are joined together by a chemical bond.
This type of block copolymers~\cite{bates90,bates91,matsen96} shows
a wealth of microphases. (ii) Side-chain liquid crystalline
polymers (LCPs). Here, liquid crystal-forming groups are linked to
polymer backbones through flexible spacers. The most prominent
phase of these systems is the smectic, where the LC groups form
orientationally ordered layers separated by disordered lamellae
containing the poymeric backbones~\cite{shibaev94,demus98}. (iii)
Ternary systems consisting of water, oil and an amphiphilic
surfactant. These systems can show a variety of microstructured
phases, with the amphiphilic surfactant stabilizing the oil-water
interfaces and thus preventing ``macrophase
separation''~\cite{gelbart94,gompper94}. All three of the cases
above are examples of thermotropic systems, i.e. systems in which
the phase behavior is governed by temperature as a controlling
variable, reflecting the dominance of energetic effects.

Recently, MPS was observed in an entirely new class of systems.
Binary mixtures of bacteriophage viruses and (small) latex spheres
with varying size ratios showed a surprisingly rich phase
behaviour, including a lamellar phase~\cite{adams98}. In this
phase, the lamellae are defined by a ``smectic'' arrangement of
the rodlike virus particles in layers with the spherical latex
particles in between the layers~\cite{adams98,adams98b}. These
results are remarkable for two reasons. First, unlike the previous
archetypal cases of MPS, we are dealing with a binary \emph{mixture}
which phase separates on a microscopic scale. There is no ``hard''
constraint like a chemical bond that prevents the two species from
phase separating on a macroscopic scale, and both species remain
in a fluid state within the layers. Second, it was argued that the
virus particles as well as the latex spheres can be modeled to a
good extent to interact as hard bodies. Consequently, the driving
force causing this MPS must be of an entropic nature. This is also
in stark contrast with MPS in block copolymers, LCP's and
amphiphiles where the dependence on temperature is strong and
hence indicates a predominantly energetic effect.
The possibility of this type of MPS was already explored in computer simulations
~\cite{koda96} and found to be qualitatively well described within
the so-called second virial approximation~\cite{koda96,dogic00}, the
validity of which can only be guaranteed at low densities.
However, as the experimental systems are far from dilute this last
treatment may not capture all the essential ingredients. It has
been argued that MPS in binary mixtures may be caused by the
so-called depletion effect~\cite{adams98,adams98b}, which is
generically a many-body interaction effect and is not well
described with a second-virial theory. Consequently, a more
accurate approach would be required, certainly in order to resolve
in detail what prevents the system from demixing macroscopically.

That entropy \emph{per se} can be the driving force for phase
transitions has by now been well established. There are many
examples ranging from ordering in monodisperse systems like the
liquid-to-crystal transition in hard spheres~\cite{alder57} and
the isotropic-to-nematic transition in slender hard
rods~\cite{onsager49}, to demixing in binary mixtures, like e.g.\
the Asakura-Oosawa (AO) mixture of hard spheres and ideal spheres,
which is used as a model for colloid-polymer
mixtures~\cite{asakura54,asakura58,vrij76}. In essence, the
physical mechanism in all these systems is the same; the gain in
effective ``free volume'' available to the particles upon ordering
offsets the loss of entropy of disorder or mixing respectively.
For the AO mixture this is usually referred to as the previously
mentioned depletion effect; the ideal polymers are
depleted from a shell around the impenetrable colloids.
Overlap of these depletion shells increases the free volume
available to the polymers and hence this system phase separates
into a colloid-rich and a colloid-poor fluid~\cite{poon02}.
However, whereas entropy can induce phase behaviour that is at
least as rich as seen in energetic systems, MPS remains a very
rare phenomenon in entropic systems~\cite{lekkerkerker98}.

A variant of the depletion effect was also recently discovered in
theoretical treatments of binary mixtures of thin and thick hard
rods~\cite{sear95,vanroij96}. These systems are seen to be able to
phase separate in two isotropic fluid phases due to depletion.
Here, however, the depletion interaction appears as a genuine two-body
effect~\cite{vanroij96}, in contrast to the AO system, in which it
is predominantly a three-body effect. Consequently, this form of
depletion effect survives the Onsager limit (length$\gg$width)
applied to both species of rods, and for sufficiently asymmetric
rods, preempts the usual transition to the orientationally ordered
nematic phase~\cite{sear95,vanroij96,sear96,bosetti01}. These
predictions have since been corroborated by
simulations~\cite{dijkstra97,vanroij98}. In the present paper, we
propose to use the two-body depletion effects between slender rods
of different diameters to construct a system which shows
entropy-induced MPS. Taking our cue from the concepts developed in
the field of thermotropic block copolymers, we connect a chain of
freely rotating ``thick'' hard rods to a chain of freely rotating
``thin'' hard rods. The above-mentioned unfavourable depletion
interaction between these two types of rods provides the tendency
to fully demix, whereas the joint (connecting the two strands)
prevents this. The so-constructed system of freely jointed hard
diblock copolymers (HDC) is in our view one of the most simple
systems conceivable showing entropy-induced MPS. Furthermore, and
contrary to the case of MPS in the binary rod-sphere mixtures, the
physical mechanism is both clear and robust. Of course, there is
as yet no direct candidate for an experimental system well
described. However, it may certainly be possible for
experimentalists to connect (possibly long and flexible)
chemically inert polymers to the ends of virus particles like
TMV~\cite{fraden94,dogic01}. Together with an appropriate solvent
this may mimick an effective rod-coil system with only hard body
interactions. In this system, the polymer tails are likely to
stabilize the smectic phase of the virus particles and this could
be viewed as a microseparated phase.

In order to describe this system we employ a density functional
theory in the second-virial or Onsager approximation starting from
first principles. We assume that multiple overlaps between two
chains as well as self-overlaps of the chains are unimportant. All
three of the above approximations, common in theoretical
treatments of
LCPs~\cite{khokhlov81,khokhlov82,vroege92,wessels03}, should
become exact in the Onsager limit where the lengths of the rods
involved is much larger than their widths. The stationarity
equations that determine the stable phases in our theory are
solved locally by means of a bifurcation (or, equivalently, linear
stability) analysis of the isotropic fluid
phase~\cite{kayser78,wessels04}. Apart from fluctuations with a
nonzero wave vector corresponding to a microseparated phase, we
also consider spatially homogeneous fluctuations with nematic
symmetry, in order to study the competition between these two
types of ordering. For both phases, we obtain closed analytical
expressions for the spinodal density. We find that for long chains
and nonzero difference in the widths, the microseparated phase
always preempts the nematic.

Naturally we want to make contact with the vast amount of
literature on thermotropic block copolymers in the weak
segregation limit. Most of these follow the original treatment
proposed in the seminal paper by Leibler~\cite{leibler80}. Leibler
considered diblock copolymers interacting via the heuristic Flory
parameter $\chi$ and constructed a Landau expansion in the average
composition fluctuations. By applying the ``random phase
approximation'' and retaining only leading orders of the Fourier
modes, he was able to map out more or less the complete phase
diagram. Subsequent refinements extended the theory to the strong
segregation regime~\cite{semenov85}, added
fluctuations~\cite{fredrickson87} and included extra
phases~\cite{matsen94}, but did not change the essence of the
approach. Leibler's results have been confirmed qualitatively by
experiments (Ref.~\cite{bates90} and Refs.\ therein) and, for
finite chains lengths~\cite{fredrickson87}, by simulations
(Refs.~\cite{fried91,micka95} and Refs.\ therein). The Leibler
approach treats the correlations within the polymers on the
Gaussian level~\cite{bates90}. We can therefore connect to this
approach by applying the Gaussian limit to our model of freely
jointed HDC's. Within this limit our theory becomes equivalent to
that of Leibler as far as the treatment of the \emph{intra}chain
interactions is concerned. However, the \emph{inter}chain
interactions between the polymers are essentially different in the
present case, as they are of a geometric nature, i.e.\ totally
fixed by the dimensions of the composing hard rods. In the
Leibler theory, these interactions are described generically by
means of the freely adjustable Flory parameter. A full exploration
of the parallels between the two approaches, however, was beyond
the scope of this work.

Another class of systems, that appears as a special case of our
model are the well-studied rod-coil diblock copolymers. These
consist of one stiff (rodlike) block and a much more flexible
part. In such systems, liquid crystalline ordering competes with
MPS and a number of theoretical studies have been devoted to the
subject. Most of these combine the Leibler approach with an
additional Maier-Saupe anisotropic orientational interaction
resulting in the appearance of a nematic phase (and sometimes an
additional smectic phase)  in the phase diagram, besides the
various microseparated
phases~\cite{semenov86,holyst92,williams93,singh94,sones94,netz96,reenders02}.
However, the ratio of the Flory and the Maier-Saupe interaction
parameters in these approaches is rather arbitrary, whereas in the
present approach microseparated and nematic ordering both result
from the same geometric origin with no room for additional tuning.

Finally, there has been some related work on more idealized but
conceptually simpler systems in the context of entropic liquid
crystals. Ho{\l}yst considered parallel nail-shaped particles which
showed a nematic-to-smectic A$_{\rm d}$ transition~\cite{holyst90}.
As a model for surfactants, Bolhuis and Frenkel studied non-additive complexes
of hard spheres and ideal spherocylinder-tails~\cite{bolhuis97a} 
where Schmidt and von Ferber
used hard slender rods for the tails~\cite{schmidt01c}
Of particular relevance to the
present work is Ref.~\cite{duechs02} where D\"{u}chs and Sullivan
investigate the phase behavior of hard \emph{wormlike} diblock
copolymers. However, in this latter work only differences in
persistence length are considered and not in thickness between the
two components.  Consequently they only find competition between
a nematic and a (orientationally ordered) smectic phase, instead
of the (orientationally \emph{dis}ordered) lamellar phase.
Moreover, only numerical solutions to the stationarity equations
are presented, whereas we are able to obtain additional analytical
insight through the stability analysis of the isotropic fluid
phase. Lastly, van Duijneveldt and Allen used Monte Carlo simulations
to study the effect of flexible tails on the phase behavior of
spherocylinders~\cite{duijneveldt97}. This was later extended by
Casey and Harrowell to rod-coil molecules of which the isolated
rods do not posses a smectic phase~\cite{casey99}.

Although our theory is formulated for chains with a finite number
of rodlike segments, we devote the major part of this paper to
chains with an infinite number of segments in which the
correlations between the segments are Gaussian. We formulate a
consistent Gaussian limit, in which the number of model parameters
reduces to just three.  The limit is chosen in such a way that we
can still consider the competition between MPS and nematic
ordering. This is achieved by letting the difference in thickness
between the two types of rods to become infinitesimally small. The
limit moreover is such that most of the assumptions in the
original derivation of the model are fully satisfied. The most
prominent results are phase diagrams as a function of the model
parameters, showing the regions of stability of the microseparated
or nematic phases. Furthermore, exploiting the features of the
bifurcation analysis, we are able to calculate the relative order
along the polymer in the microseparated phase at the bifurcation
point. The outline of the paper is as follows: in Sec.~II we
define the model and develop the formalism. In Sec.~III we briefly
discuss the symmetry of the phases involved. The bifurcation
analysis is the topic of Sec.~IV and the Gaussian limit is applied
in Sec.~V. Sec.~VI is the results section and we end with a
discussion in Sec.~VII.

\section{Model and Formalism}
\label{sec:model}

\begin{figure}[t]
\bp(12,2.5)
\put(0,0){\epsfig{figure=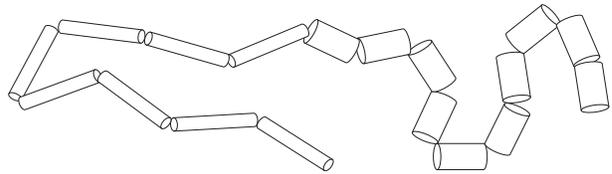,width=8cm}}
\ep
\caption{\small An example of a hard diblock copolymer.
A freely-jointed chain of $M_{\rm A}$ hard rods with dimensions $l_{\rm A}$ and $d_{\rm A}$ (left side)
are connected to a freely-jointed chain of $M_{\rm B}$ hard rods with dimensions 
$l_{\rm B}$ and $d_{\rm B}$ (right side).}
\label{fig:polymer}
\end{figure}

We consider a monodisperse fluid of $N$ diblock copolymers in a
volume $V$. Each polymer is a chain of freely-jointed cylindrical
rods connected end-to-end where the first $M_{\rm A}$ rods are of
type $\tau=A$ having length $l_{\rm A}$ and width $d_{\rm A}$ and
the last $M_{\rm B}$ rods are of type B with dimensions $l_{\rm
B}$ and $d_{\rm B}$ (see Fig.~\ref{fig:polymer}). We assume that both types of rods are very
slender, $l_{\tau}\gg d_{\tau}$, with $\tau\in\{{\rm A,B}\}$,
hard bodies, i.e.\ impenetrable to other rods. The total number of
segments in a chain is $M=M_{\rm A}+M_{\rm B}$ and every segment
has a label $m\in\{1,\ldots,M\}$ specifying its position in the
chain. The state of a segment is described by the position ${\bf
r}_m$ of its center of mass and an orientation, being a unit
vector $\ho_m$ pointing along its long axis in the direction of
increasing $m$. The configuration of a whole chain $\bfm{\xi}$ is
fully characterized by the position of one of its segments (say
the first; ${\bf r}_1$) and the orientations of all of them,
$\bfm{\Omega}=\{\ho_1,\ldots,\ho_{M}\}$, so $\bfm{\xi}=\{{\bf
r}_1,\bfm{\Omega}\}$. The position of a segment $m$ is then given
by ${\bf r}_m={\bf
r}_1+\frac{1}{2}\sum_{k=1}^{m-1}(l_k\ho_k+l_{k+1}\ho_{k+1})$ where
$l_k=l_{\rm A}$ if $k\leq M_{\rm A}$ and $l_{\rm B}$ if $k\geq
M_{\rm A}+1$.

In density functional theory (DFT) the free energy of a (possibly inhomogeneous)
fluid of molecules is expressed as a functional
of the single-molecule configuration distribution function, $\rho^{(1)}(\bfm{\xi})$~\cite{evans92}.
Using the second-virial (or Onsager) approximation
it is formulated as follows~\cite{allen93}
\begin{multline}
\beta{\cal F}\left[\rho^{(1)}\right]
=\int d\bfm{\xi}
\rho^{(1)}(\bfm{\xi})
\left[
\log \left( {\cal V}_{\rm T} \rho^{(1)}(\bfm{\xi}) \right)-1
\right] \\
-{\textstyle \frac{1}{2}}
\int\int d\bfm{\xi}d\bfm{\xi}'\rho^{(1)}(\bfm{\xi})\rho^{(1)}(\bfm{\xi}')
\Phi (\bfm{\xi},\bfm{\xi}').
\label{eq:freeenfunc1}
\end{multline}
The integrals are over single-molecule configuration space where $\int d\bfm{\xi}=
\int d{\bf r}_0d\bfm{\Omega}$ and $\int d\bfm{\Omega}=\int\prod_{m}d\ho_m$ and
$\int d\ho=\int_0^{2\pi}d\phi\int_0^{\pi}d\theta \sin\theta$.
Further, $\rho^{(1)}(\bfm{\xi})$ is normalized as
follows $\int\rho^{(1)}(\bfm{\xi})d\bfm{\xi}=N$.
The factor $\beta$ equals $(k_B T)^{-1}$ in which $k_B$ is Boltzmann's
constant and $T$ the temperature.
The volume ${\cal V}_{\rm T}$ we call the `thermal volume' and is
a product of the de Broglie thermal wavelengths of the molecules~\cite{hansen86,allen93}.
The quantity $\Phi (\bfm{\xi},\bfm{\xi}')$ is the Mayer function
of two molecules with configurations $\bfm{\xi}$ and $\bfm{\xi}'$.
As we are dealing with hard segments, the potential energy
$v(\bfm{\xi},\bfm{\xi}')$ between two chains is
$\infty$ when they overlap and 0 when they don't.
Consequently, the Mayer function is given by
\be
\Phi (\bfm{\xi},\bfm{\xi}')=
\exp \left(-\beta v(\bfm{\xi},\bfm{\xi}')\right)-1=
\left\{
\ba{rl}
-1&\mbox{if overlap}\\
0&\mbox{if no overlap} \ea . \right. \label{eq:mayer} \ee The
configurations of both chains involved can be highly irregular and
the dependence of $\Phi$ very complicated. Therefore we
approximate the chain-chain Mayer function $\Phi$ by the sum of
all the segment-segment Mayer functions $\phi_{m,m'}$, \be \Phi
(\bfm{\xi},\bfm{\xi}')=\sum_{m,m'=1}^{M} \phi_{m,m'} ({\bf
r}_{m}-{\bf r}_{m'},\ho_m,\ho_{m'}). \label{eq:mayer2} \ee This
expression, to which only individual pairs of segments contribute,
is actually the first term in a systematic expansion of the Mayer
function. Higher order terms involve interactions between more
than two segments simultaneously~\cite{wessels03}. At this point
we note that apart from (i) the second virial approximation and
(ii) the above expression for the chain-chain Mayer function,
another (iii) approximation has been made. In this formalism the
chains are allowed to self overlap, i.e. other than the spatial
constraint that successive segments are connected to each other
there are no interactions within the chain. All three of these
approximations are commonly used and corrections to the first two
are small when $l_{\tau}\gg d_{\tau}$~\cite{khokhlov81,khokhlov82,vroege92}. 
The neglect
of the effects of self-overlap is assumed to be reasonable in a
dense polymer melt~\cite{degennes79} where screening effectively
compensates the intramolecular interactions and as a result interactions
between distant parts of the same chain are indistinguishable from
interactions with the average environment because of loss of
intrachain correlations.

In thermodynamic equilibrium, the free energy reaches a minimum and
the functional is stationary.
Therefore, we consider the variation of Eq.~\ref{eq:freeenfunc1}
with respect to $\rho^{(1)}$,
\be
\frac{\delta}{\delta \rho^{(1)}(\bfm{\xi})} \beta{\cal F}
-\beta \mu=0
\label{variation}
\ee
with the chemical potential $\mu$ playing the role of Lagrange
multiplier needed to enforce normalization.
Eliminating $\mu$ from Eq.~\ref{variation} yields the
(selfconsistent) stationarity equation,
\be
\rho^{(1)}(\bfm{\xi})=\frac{
N \exp\left[
\int d\bfm{\xi}' \rho^{(1)}(\bfm{\xi}')
\Phi_M (\bfm{\xi},\bfm{\xi}')
\right]}{ \int d\bfm{\xi}
\exp\left[
\int d\bfm{\xi}' \rho^{(1)}(\bfm{\xi}')
\Phi_M (\bfm{\xi},\bfm{\xi}')
\right]},
\label{eq:stationarity1}
\ee

In order to proceed, we define the single-segment distribution function (SDF)
(of segment $m$), $\rho_m({\bf r}_m,\ho_m)$, in the following way
\begin{multline}
\rho_m ({\bf r}_m,\ho_m)=\int \prod_{k\neq m}d\ho_k  \rho^{(1)}(\bfm{\xi})
\\ =\int \prod_{k\neq m}d\ho_k
\rho^{(1)} ({\bf r}_1({\bf r}_m,\bfm{\Omega}),\bfm{\Omega}),
\label{eq:projection}
\end{multline}
in which ${\bf r}_1$ is given by
${\bf r}_m-\frac{1}{2}\sum_{k=1}^{m-1}(l_k\ho_k+l_{k+1}\ho_{k+1})$
and the product is over all segments $k$ but the $m$th.
Integrating Eq.~\ref{eq:stationarity1} over all $\ho_k$
except for $\ho_m$ as well and using Eq.~\ref{eq:mayer2}
we obtain a set of equations,
\begin{multline}
\label{eq:stationarity1p}
\rho_m({\bf r}_m,\ho_m)=
\frac{N}{Q} \int  \prod_{k\neq m}d\ho_k \times \\ \exp\left[
\sum_{k,k'=1}^M\int d{\bf r}'_{k'}d\ho'_{k'} \rho_{k'}({\bf r}'_{k'},\ho'_{k'})
\phi_{k,k'} ({\bf r}_{k}-{\bf r}'_{k'},\ho_k,\ho'_{k'})
\right],
\end{multline}
where $Q$ is the normalization factor; i.e.\ the SDF's are
normalized in the same way as $\rho^{(1)}$: i.e.\
$\int d{\bf r}d\ho \rho_m({\bf r},\ho)=N$.

\section{Phase Behaviour and Order Parameters}
\label{sec:phases}

\subsection{Isotropic Phase}
\label{sub:isotropic}

At low polymer number density, $n=N/V$, the system is in the isotropic
fluid phase, and $\rho_{m}({\bf r}_{m},\ho_{m})$ is a constant, so due to normalization,
$\rho_{m}^{\rm (iso)}=n/4\pi$.
Consequently,
\begin{align}
\nonumber
\int d{\bf r}' d\ho' \rho_{m}^{\rm (iso)}  & \phi_{k,k'} ({\bf r}_{k}-{\bf r}',\ho_k,\ho')
\\ &=-\frac{n}{4\pi}\int d\ho'  l_kl_{k'}(d_k+d_{k'})\sin\gamma(\ho_k\cdot\ho') \\
&=-{\textstyle \frac{1}{4}} \pi n l_kl_{k'}(d_k+d_{k'}), \nonumber
\end{align}
where $\gamma (\ho\cdot\ho')$ is the planar angle between $\ho$ and $\ho'$
and one can recognize $l_kl_{k'}(d_k+d_{k'})\sin\gamma(\ho,\ho')$
as the excluded volume of two rods $k$ and $k'$ with respective
orientations $\ho$ and $\ho'$.
This yields the following normalization factor in the isotropic phase
\begin{multline}
Q_{\rm iso}=(4\pi)^{M_{\rm A}+M_{\rm B}}V \times \\
\exp \left[
-{\textstyle \frac{1}{2}} \pi n\left(
M_{\rm A}^2l_{\rm A}^2d_{\rm A}+M_{\rm A}M_{\rm B}l_{\rm A}l_{\rm B}(d_{\rm A}+d_{\rm B})
+M_{\rm B}^2l_{\rm B}^2d_{\rm B}
\right)\right]
\label{eq:isotropic}
\end{multline}
Choosing the dimensions of rod A as units, we define
\be
\ba{cccc}
\tilde{l}=l_{\rm B}/l_{\rm A},&\tilde{d}=d_{\rm B}/d_{\rm A},&\tilde{M}=M_{\rm B}/M_{\rm A},
\ea
\ee
and a dimensionless segment density in a symmetric way,
\be
\tilde{n}=2n(M_{\rm A}l^2_{\rm A}d_{\rm A}+M_{\rm B}l^2_{\rm B}d_{\rm B}).
\ee
Then, Eq.~\ref{eq:isotropic} becomes
\be
Q_{\rm iso}=(4\pi)^MV
\exp \left[
-\frac{\pi}{4} \tilde{n}M
\frac{(1+\tilde{M}\tilde{l})(1+\tilde{M}\tilde{l}\tilde{d})}{(1+\tilde{M})(1+\tilde{M}\tilde{l}^2\tilde{d})}
\right],
\label{eq:isotropic2}
\ee
where we have also used $M=M_{\rm A}+M_{\rm B}$.
We also note that the normalization factor $Q_{\rm iso}$ is exactly the partition sum of the block copolymers
in the isotropic phase.

\subsection{Nematic Phase}
\label{sub:nematic}

In the (uniaxial) nematic phase, there is orientational order
with respect to a direction $\hat{n}$, however,
the system is still spatially homogeneous.
Therefore, the SDF can be expanded in Legendre polynomials.
\be
\rho_m({\bf r},\ho)=\rho_m(\ho)=n\sum_{j=0}^{\infty}\frac{2j+1}{4\pi}
a_m^{(j)}P_j(\ho\cdot\hat{n}),
\ee
with coefficients
\be
n a_m^{(j)}=\int d\ho'P_j(\ho'\cdot\hat{n})\rho_m(\ho').
\ee
Due to normalization, $a_m^{(0)}=1$ as it is in the isotropic phase
and due to up-down symmetry of the nematic, all $a_m^{(j)}=0$ for odd $j$ (in the isotropic
fluid, $a_m^{(j)}=0$ for all $j\neq 0$).
The lowest-order coefficient different in the nematic and the
isotropic phase is $a_m^{(2)}$ which is the usual Maier-Saupe order parameter.
The physical incentive to form a nematic is that the average excluded volume between
rods is smaller (and therefore the average free volume available to the rods is larger)
in the nematic phase~\cite{onsager49}.

\subsection{Microseparated Phase}
\label{sub:micro}

Microseparated phases consist of spatially distributed regions
rich either in type-A  or type-B rods and are typically governed
by a single dominant wavelength. These phases exist in a variety
of types exhibiting various degrees of symmetry, e.g.\ lamellar,
hexagonal, bcc and even more exotic morphologies like the gyroid
phase~\cite{bates90,matsen94}. In this paper we do not consider
the various symmetries of microseparated phases but focus on the
magnitude of the dominant wavelength and the density for which it
becomes unstable. To that end, we observe that the SDF can be
expanded in terms of plane waves \be \rho_m ({\bf r},\ho)=
\sum_{{\bf q}\in\mathbb{L}} \hat{\rho}_m ({\bf q},\ho) {\rm e}^{i
{\bf q}\cdot{\bf r}} \ee with $\mathbb{L}$ some set of wave
vectors and the ``coefficients'' given by \be \hat{\rho}_m ({\bf
q},\ho) = V^{-1}\int d{\bf r}'{\rm e}^{-i {\bf q}\cdot{\bf r}'}
\rho_m ({\bf r}',\ho). \ee In general there will be orientational
order within the domains and consequently the coefficients still
depend on the orientation. If needed, one could proceed and expand
these coefficients again in spherical harmonics. However, in order
to simplify the analysis, this additional orientational order in
the microseparated phase is usually neglected which we will show
in Sec.~\ref{sec:hdgl} is permitted for the case of infinitely
long polymers. In homogeneous fluid phases like the nematic, the
SDF is independent on the spatial coordinate and only the
coefficient $\hat{\rho}_m ({\bf 0},\ho)$ at zero wavenumber
survives.

\section{Bifurcation Analysis}
\label{sec:bifan}

\subsection{The Bifurcation Equation}
\label{sub:bifeq}

At low densities, the isotropic phase is the globally stable
phase, but at higher densities it will become unstable with
respect to lower symmetry phases exhibiting some form of ordering.
Points where these lower-symmetry solutions branch off the
isotropic solution are called bifurcation points and the densities
at which this happens, bifurcation densities. Different solutions
may bifurcate at different densities from the isotropic phase.
Generically the particular solution which bifurcates at the lowest
density, will give rise to the first ordered phase that is also
thermodynamically stable with respect to the isotropic phase. In
this section, we perform a linear stability (or bifurcation)
analysis around the isotropic parent solution, along the lines of
Refs.~\cite{kayser78,wessels03,wessels04}. Consequently, we assume
isotropic distributions with a perturbation of lower symmetry, \be
\rho_m({\bf r},\ho)= \frac{n}{4\pi}+ \varepsilon \rho_{m,1}({\bf
r},\ho), \ee where the proper normalisation of the SDF requires
$\int d{\bf r}d\ho \rho_{m,1}({\bf r},\ho)=0$. Inserting this in
the stationarity equations~\ref{eq:stationarity1p} we linearize
the exponent with respect to the infinitesimal parameter
$\varepsilon$,
\begin{multline}
\exp\left[
\sum_{k,k'=1}^M\int d{\bf r}'d\ho' \rho_{k'}({\bf r}',\ho')
\phi_{k,k'} ({\bf r}_{k}-{\bf r}',\ho_k,\ho')
\right]= \\
\exp\left[-{\textstyle \frac{1}{4}} \pi n \sum_{k,k'=1}^M l_kl_{k'}(d_k+d_{k'})
\right]\times \\ \left(1+\varepsilon \sum_{k,k'}
\int d{\bf r}' d\ho' \rho_{k',1}({\bf r}',\ho')
\phi_{k,k'} ({\bf r}_k-{\bf r}',\ho_k,\ho')\right).
\end{multline}
Equating orders in $\varepsilon$, to zeroth order, we re-obtain
the isotropic result, Eq.~\ref{eq:isotropic}. To first order this
yields the so-called bifurcation equations, 
\begin{multline}
\rho_{m,1}({\bf
r}_m,\ho_m)= \frac{n}{(4\pi)^M}\int \prod_{k''\neq m} d\ho_{k''}\times \\
\sum_{k,k'} \int d{\bf r}' d\ho' \rho_{k',1}({\bf r}',\ho')
\phi_{k,k'} ({\bf r}_k-{\bf r}',\ho_k,\ho').
\label{eq:bifurcation} 
\end{multline} 
These can be interpreted a generalized
linear eigenvalue problem with eigenfunctions $\rho_{m,1}({\bf
r},\ho)$ and eigenvalue $n$, the bifurcation density. There is an
infinite hierarchy of solutions to Eq.~\ref{eq:bifurcation} for
varying degrees of symmetry. However, we are only interested in
the one (or the few) corresponding to the lowest bifurcation
density. Note that the explicit dependence on the normalization
factor $Q$ has dropped out since integration over ${\bf r}_m$ and
$\ho_m$ trivially yields zero on the left hand side by definition
and, after rearrangement of the integrals made possible by the
finite range of the Mayer functions $\phi_{k,k'}$, also on the
right hand side.

In order to make the bifurcation equation,
Eq.~\ref{eq:bifurcation}, more transparent we define for the
moment as an auxiliary quantity, the fields, \be H_k({\bf
r}_k,\ho_k)=\sum_{k'} \int d{\bf r}' d\ho' \rho_{k',1}({\bf
r}',\ho') \phi_{k,k'} ({\bf r}_k-{\bf r}',\ho_k,\ho').
\label{eq:auxiliary} \ee in terms of which the bifurcation equation becomes
\be \rho_{m,1}({\bf r}_m,\ho_m)= \frac{n}{(4\pi)^M}\sum_{k} \int
\prod_{k''\neq m} d\ho_{k''} H_k({\bf r}_k,\ho_k)
\label{eq:bifurcationp} \ee However, this field $H_k$ is a
function of ${\bf r}_k$ and $\ho_k$ whereas on the left of
Eq.~\ref{eq:bifurcationp} we have a function of ${\bf r}_m$ and
$\ho_m$. And these are not independent and as \be {\bf r}_m={\bf
r}_k +{\cal P}_{k,m} \label{eq:constraint} \ee where the vectorial
``path'' ${\cal P}_{k,m}$ between $k$ and $m$ is given by \be
{\cal P}_{k,m}={\textstyle \frac{1}{2}}\sum_{k'=k}^{m-1} \left(
l_{k'}\ho_{k'}+l_{k'+1}\ho_{k'+1} \right), \label{eq:path} \ee for
$k<m$. Further, ${\cal P}_{m,m}=0$ and the case of $k>m$ can be
obtained by realizing that ${\cal P}_{k,m}=-{\cal P}_{m,k}$.
Consequently, the interlying orientional integrations in
Eq.~\ref{eq:bifurcationp} have to make the connection and
``transfer'' the field from segments $k$ to $m$.

We return to Eq.~\ref{eq:bifurcation} and insert the constraint,
Eq.~\ref{eq:constraint} via a delta function
\begin{multline}
\label{eq:bifurcation2}
\rho_{m,1}({\bf r}_m,\ho_m)=\\
\frac{n}{(4\pi)^M} \sum_k \int \prod_{k''\neq m}d\ho_{k''} 
\int d{\bf r}_k \delta ({\bf r}_m-{\bf r}_k +{\cal P}_{m,k})\times \\
\sum_{k'}
\int d{\bf r}' d\ho' \rho_{k',1}({\bf r}',\ho')
\phi_{k,k'} ({\bf r}_k-{\bf r}',\ho_k,\ho').
\end{multline}
Next, we observe that in Eq.\ref{eq:bifurcation2} there appear two
spatial convolution integrals.
Therefore, it makes sense to proceed with a Fourier transform
(i.e. $\hat{g}({\bf q})=V^{-1}\int d{\bf r}_m {\rm e}^{-i {\bf q}
\cdot {\bf r}_m} g({\bf r}_m)$), yielding
\begin{multline}
\label{eq:bifurcation3}
\hat{\rho}_{m,1}({\bf q},\ho_m)=
\frac{n}{(4\pi)^M} \sum_k \int \prod_{k''\neq m}d\ho_{k''}
{\rm e}^{-i {\bf q}\cdot {\cal P}_{k,m}}\times \\
\sum_{k'}
\int d\ho' \hat{\rho}_{k',1}({\bf q},\ho')
\hat{\phi}_{k,k'} ({\bf q},\ho_k,\ho').
\end{multline}
This is the general form of the bifurcation equation
for a lower-symmetry solution bifurcating off the isotropic-fluid parent solution.
Note that the ${\bf q}$-vector is the same for all segments.
Furthermore, at this point, we have not yet specified the internal structure of
the polymer, only that it is a chain of cylindrically-symmetric (rodlike) segments which contains
no closed loops.
Concerning the rodlike segments, the Fourier transformed Mayer function 
$\hat{\phi}_{k,k'}$ is calculated in
Appendix~\ref{sec:ftmayer} and is for very slender segments ($l_k\gg d_k$) given by,
\begin{multline}
\hat{\phi}_{k,k'}({\bf q},\ho_k,\ho_{k'})=-
l_{k}l_{k'}(d_{k}+d_{k'})|\ho_{k}\times\ho'_{k'}| \times \\
j_0\left({\textstyle \frac{1}{2}}l_k {\bf q}\cdot\ho_k\right)
j_0\left({\textstyle \frac{1}{2}}l_{k'} {\bf q}\cdot\ho_{k'}\right),
\label{eq:Fmayer}
\end{multline}
where we have already discarded higher-order terms
containing $(d_{k}+d_{k'}){\bf q}$ as the wave vector will be at most of
order $1/l_{\rm A,B}$ so these terms will be small.
The function $j_0(x)=\sin x/x$ is the spherical Bessel function of zeroth order.
We proceed by solving Eq.~\ref{eq:bifurcation3} to which we refer as {\em the} bifurcation
equation from now on.

\subsection{Nematic Solution}
\label{sub:nematicsolution}

We first consider the nematic solution, which is also the
simplest. In the nematic phase, ${\bf q}=0$ and the orientational
integrals in the bifurcation equation are trivial and it reduces
to \be\label{eq:bifurcationnem} \hat{\rho}_{m,1}(\ho_m)=
\frac{n}{4\pi}\sum_{k'} \int d\ho' \hat{\rho}_{k',1}(\ho')
\hat{\phi}_{m,k'}(\ho_m,\ho'), \ee where
$\hat{\rho}_{m,1}(\ho_m)=\hat{\rho}_{m,1}({\bf 0},\ho_m)$ and \be
\hat{\phi}_{m,k'}(\ho_m,\ho_{k'})=
-l_{m}l_{k'}(d_{m}+d_{k'})|\ho_{m}\times\ho'_{k'}|.
\label{eq:minusexclvol} \ee is simply minus the excluded volume of
two rods with fixed orientations, $\ho_m$ and $\ho_{k'}$. This
bifurcation equation is the same as that of a mixture {\em
disconnected} rods~\cite{vanroij96}, so for orientational ordering
the connectivity of the rods within the chains does not play a
role. The kernel $\hat{\phi}_{m,k'}$ is now only a function of the
planar angle $\gamma$ between the orientations of the rods,
$|\ho_{m}\times\ho'_{k'}|= |\sin\gamma(\ho_m\cdot\ho_{k'})|$.
Consequently, due to this uniaxial symmetry the eigenfunctions of
$\hat{\phi}_{m,k'}$ and therefore of Eq.~\ref{eq:bifurcationnem}
are simply the Legendre polynomials $P_j$ (see
Appendix~\ref{sec:eigenfunctions}), \be \int d\ho'
\hat{\phi}_{m,k'}(\ho,\ho')P_j(\ho'\cdot\hat{n})=
-l_ml_{k'}(d_m+d_{k'})s_j P_j(\ho\cdot\hat{n}), \ee with $s_j$ the
Legendre coefficients of $|\sin\gamma|$. In case of the nematic
phase, it is well known that this becomes first unstable with
respect to the mode $j=2$, so $\hat{\rho}_{m,1}(\ho_m)=(5/4\pi) n
c_m^{(2)}P_2(\ho_m\cdot\hat{n})$ with $c_m^{(2)}$ the Legendre
coefficients. Then, the bifurcation equation becomes \be
\label{eq:bifurcationnem2} c_m^{(2)}=
-\frac{n}{4\pi}\sum_{k'}l_ml_{k'}(d_m+d_{k'})s_2 c_{k'}^{(2)} \ee
with $s_2=-\pi^2/8$. This is an $M\times M$ matrix eigenvalue
equation and therefore in principle much too large to solve.
However, by observing that the geometric factor on the right hand
side does not so much depend on the segments $m,k'$ but on whether
they belong to A or B, we can split the summation,
$\sum_{k'}=\sum_{\tau}\sum_{k'\in\tau}$ with $\tau ={\rm A,B}$.
Then, we can define the ``type-average'' coefficients,
$c_{\tau}^{(2)}=(1/M_{\tau})\sum_{m\in\tau}c_{m}^{(2)}$ and
Eq.~\ref{eq:bifurcationnem2} becomes, \be
\label{eq:bifurcationnem3} c_{\tau}^{(2)}= \frac{\pi
n}{32}\sum_{\tau'={\rm A,B}} M_{\tau'}
l_{\tau}l_{\tau'}(d_{\tau}+d_{\tau'}) c_{\tau'}^{(2)}. \ee
Rewriting this in terms of dimensionless quantities, \be
\label{eq:bifurcationnem4} {\bf c}_2= {\displaystyle \frac{\pi
\tilde{n}}{32(1+\tilde{M}\tilde{l}^2\tilde{d})}}{\bf G}_2 {\bf
c}_2 \ee with \be\ba{ccc} \label{eq:bifurcationnem4p}
{\bf G}_2=\left[\ba{cc}1 &{\textstyle \frac{1}{2}}\tilde{M}\tilde{l}(1+\tilde{d})\\
{\textstyle \frac{1}{2}}\tilde{l}(1+\tilde{d})&
\tilde{M}\tilde{l}^2\tilde{d} \ea\right] &\mbox{and}& {\bf
c}_2=\left(\ba{c}c_{\rm A}^{(2)}\\ c_{\rm B}^{(2)}\ea\right)
\ea\ee we now have reduced the problem to a simple $2\times 2$
matrix eigenvalue equation. There are two solutions for the
density, 
\begin{multline}
\tilde{n}_{\pm}= \frac{32(1+\tilde{M}\tilde{l}^2\tilde{d})}{\pi}
\times \\ \left(\tr{\bf G}_2\pm\sqrt{\tr^2{\bf G}_2-4\det{\bf G}_2}
\right)/(2 \det{\bf G}_2), \label{eq:eigenvaluenem} 
\end{multline} 
with
$\det$ and $\tr$ denoting the determinant and trace respectively.
As the determinant of ${\bf G}_2$ is negative, only the minus sign
in Eq.~\ref{eq:eigenvaluenem} yields a positive bifurcation
density $\tilde{n}_{\rm nem}$, so \begin{multline} \tilde{n}_{\rm
nem}=\frac{32(1+\tilde{M}\tilde{l}^2\tilde{d})}{\pi}\times \\ \left(\tr{\bf
G}_2-\sqrt{\tr^2{\bf G}_2-4\det{\bf G}_2} \right)/(2 \det{\bf
G}_2). \label{eq:eigenvaluenem2} \end{multline}  Note that, within the context
of the model as introduced in Sec.~\ref{sec:model}, this analytic
expression for the nematic bifurcation is an exact result. In the
wider context of liquid crystalline polymers, a more general
derivation of the nematic bifurcation density can be found in
Ref.~\cite{wessels04}.

\subsection{Microseparated Solution}
\label{sub:mspsolution}

In a microseparated phase, the wave vector ${\bf q}$ is nonzero and
the orientational integrals in the bifurcation equation
have to be performed explicitly.
However, we can make much progress by observing that most of the integrals
are still trivial, i.e.\ if segment $k''$ does not lie between $k$ and $m$ it does not help
to ``pass on'' the infinitesimal field $H_k$ or equivalently, there is no dependence in
the factor $\exp(-i {\bf q}\cdot {\cal P}_{k,m})$.
Consequently, these $M-|m-k|-1$ integrations each contribute a factor $\int d\ho=4\pi$
which is in total $(4\pi)^{M-|m-k|-1}$.
On the other hand, concerning the intermediate segments $k''$
between $k$ and $m$; the only dependence on $\ho_{k''}$ is in the path
${\cal P}_{k,m}$.
Therefore, suppose for a moment that $k+1<m$,
\begin{multline} 
\int \prod_{k''=k+1}^{m-1}d\ho_{k''}
{\rm e}^{-i {\bf q}\cdot {\cal P}_{k,m}}= \\
{\rm e}^{-\frac{1}{2}i {\bf q}\cdot l_k\ho_k}
\left( \prod_{k''=k+1}^{m-1} \int d\ho
{\rm e}^{-i {\bf q}\cdot l_{k''}\ho}\right)
{\rm e}^{-\frac{1}{2}i {\bf q}\cdot l_m\ho_m},
\label{eq:pathintegration}
\end{multline} 
and it is easy to show that
\be
\int d\ho {\rm e}^{- i {\bf q}\cdot l_{k''} \ho}=
4\pi\frac{\sin ql_{k''}}{ql_{k''}}=4\pi j_0(ql_{k''}),
\label{eq:dummyfactor}
\ee
where we have used ${\bf q}=q\hat{q}$ with $q$ being the length and
the unit vector $\hat{q}$ the direction of the wave vector.
When $m+1<k$, there is an extra minus sign as ${\cal P}_{m,k}=-{\cal P}_{k,m}$
but this does not change the result~\ref{eq:dummyfactor}, only the end
factors in Eq.~\ref{eq:pathintegration}.
Consequently, we define the factor
\be
F_{k,m}(q)=\left\{\ba{lll}
\prod_{k''=k+1}^{m-1} j_0(ql_{k''})
&\mbox{for}&k<m-1\\
1&\mbox{for}&k=m-1,m\\
\ea\right. ,
\label{eq:defF}
\ee
which is symmetric so $F_{k,m}(q)=F_{m,k}(q)$.
Inserting this in the bifurcation equation yields,
\begin{multline}
\hat{\rho}_{m,1}({\bf q},\ho_m)=\\
\frac{n}{4\pi}\sum_{k'}
\int d\ho' \hat{\rho}_{k',1}({\bf q},\ho')
\hat{\phi}_{m,k'} ({\bf q},\ho_m,\ho')+ \\
\frac{n}{(4\pi)^2}
\sum_{k\neq m} {\rm e}^{\sigma_{m,k}\frac{1}{2}i {\bf q}\cdot l_m\ho_m}
F_{m,k}(q)\sum_{k'}\times \\
\int d\ho d\ho' \cos\left({\textstyle \frac{1}{2}} {\bf q}\cdot l_k\ho\right)
\hat{\rho}_{k',1}({\bf q},\ho')
\hat{\phi}_{k,k'} ({\bf q},\ho,\ho'),
\label{eq:bifurcation4}
\end{multline}
where $\sigma_{k,m}={\rm sign}(m-k)$ is the sign of $m-k$. Instead
of the other ``end factor'' $\exp\left(\sigma_{m,k}{\textstyle
\frac{1}{2}}i {\bf q}\cdot l_k\ho\right)$ we have used
$\cos\left({\textstyle \frac{1}{2}} {\bf q}\cdot l_k\ho\right)$ as
within the integral only the even part in ${\bf q}$ survives. The
first term on the right hand side is due to the infinitesimal
field $H_m$ directly on segment $m$; the second term contains the
contributions $H_k$ on segments $k\neq m$ which are being
transferred to segment $m$ via $F_{m,k}$. At this point we note
that it is impossible to solve Eq.~\ref{eq:bifurcation4}
analytically for general ${\bf q}$ and we will introduce an
approximation justified for very long polymers, $M_{\rm A},M_{\rm
B}\gg 1$. In this case the relevant wave vector is expected to be
small in magnitude and consequently, the ``end factors'' as well
as the wave dependence of $\hat{\phi}_{k,k'}$ are negligible.
Therefore, we replace them by their zeroth order approximations in
${\bf q}$, \be \hat{\phi}_{k,k'}=-
l_{k}l_{k'}(d_{k}+d_{k'})|\ho_{k}\times\ho'_{k'}|
\label{eq:approx1} \ee and
\begin{align}
&\exp\left(\sigma_{m,k}{\textstyle \frac{1}{2}}i {\bf q}\cdot l_m\ho_m\right)=1 \nonumber\\
&\cos\left({\textstyle \frac{1}{2}}i {\bf q}\cdot l_k\ho_k\right)=1
\label{eq:approx23}
\end{align}
Then the bifurcation equation becomes
\begin{multline}
\hat{\rho}_{m,1}({\bf q},\ho_m)=
\frac{n}{4\pi}\sum_{k'}
\int d\ho' \hat{\rho}_{k',1}({\bf q},\ho')
\hat{\phi}_{m,k'} (\ho_m,\ho')+ \\
\frac{n}{(4\pi)^2}
\sum_{k\neq m} F_{m,k}(q)\sum_{k'}
\int d\ho d\ho'
\hat{\rho}_{k',1}({\bf q},\ho')
\hat{\phi}_{k,k'} (\ho,\ho'),
\label{eq:bifurcation5}
\end{multline}
where again as in the case of the nematic solutions, $\hat{\phi}_{k,k'} (\ho,\ho')$
has the convenient property that it maps $P_j$ on $P_j$.
Then the only mode for which the second term on the right hand side of Eq.~\ref{eq:bifurcation5}
survives (and we have wave dependence) is for $P_0$. (For $j\neq 0$ we simply re-obtain the nematic
bifurcation equation, Eq.~\ref{eq:bifurcationnem2}.)
Consequently, integrating both sides over $\ho_m$, we obtain
\be
c_m^{(0)}(q)=-\frac{n}{4\pi}
\sum_{k} F_{m,k}(q)\sum_{k'}
l_kl_{k'}(d_k+d_{k'})s_0
c_{k'}^{(0)}(q)
\label{eq:bifurcation6}
\ee
where we have defined $c_m^{(0)}(q)=\int d\ho_m \hat{\rho}_{m,1}({\bf q},\ho_m)$ and where
\be
\int d\ho\hat{\phi}_{k,k'} (\ho,\ho')=-l_kl_{k'}(d_k+d_{k'})s_0,
\ee
with $s_0=\pi^2$.
The rest of the analysis is similar to the nematic case: again we have an $M\times M$ eigenvalue
equation and we make use of the property of the geometric factor that it depends on
the types involved and not on the segment labels, hence $\sum_{k'}=\sum_{\tau'}\sum_{k'\in\tau'}$
with $\tau'={\rm A,B}$.
Defining $c_{\tau}^{(0)}(q)=(1/M_{\tau})\sum_{k\in\tau} c_k^{(0)}(q)$
and $F_{\tau,\tau'}=(1/M_{\tau}M_{\tau'})\sum_{m\in\tau}\sum_{k'\in\tau'}F_{m,k}$,
Eq.~\ref{eq:bifurcation6} becomes
\begin{multline} 
c_{\tau}^{(0)}(q)=\\ -\frac{\pi n}{4}
\sum_{\tau'} F_{\tau,\tau'}(q)\sum_{\tau''}
M_{\tau'}M_{\tau''}l_{\tau'}l_{\tau''}(d_{\tau'}+d_{\tau''})
c_{\tau''}^{(0)}(q)
\label{eq:bifurcation7}
\end{multline} 
Rewriting in terms of dimensionless quantities, we obtain
\be
{\bf c}_0(q)=-{\displaystyle \frac{\pi\tilde{n}M}{4(1+\tilde{M})(1+\tilde{M}\tilde{l}^2\tilde{d})}}{\bf F}(q){\bf G}_0{\bf c}_0(q)
\ee
with
\begin{multline} 
{\bf G}_0=\left[\ba{cc}1 &{\textstyle \frac{1}{2}}\tilde{M}\tilde{l}(1+\tilde{d})\\
{\textstyle \frac{1}{2}}\tilde{M}\tilde{l}(1+\tilde{d})&
\tilde{M}^2\tilde{l}^2\tilde{d}
\ea\right]\\ \mbox{and} \quad {\bf c}_0(q)=\left(\ba{cc}c_{\rm A}^{(0)}\\ c_{\rm B}^{(0)} \ea\right)(q).
\end{multline}
The elements of ${\bf F}(q)$ are
\begin{multline} 
F_{\rm A,A}=
\frac{1}{M_{\rm A}^2}\left(M_{\rm A}+ \right.\\ \left.
\frac{2}{1-j_0(ql_{\rm A})}
\left\{
(M_{\rm A}-1)-\frac{j_0(ql_{\rm A})-(j_0(ql_{\rm A}))^{M_{\rm A}}}{1-j_0(ql_{\rm A})}
\right\}\right),
\nonumber
\end{multline} 
\begin{multline} 
F_{\rm A,B}=F_{\rm B,A}=
\left(\frac{1}{M_{\rm A}}\frac{
1-(j_0(ql_{\rm A}))^{M_{\rm A}}}{1-j_0(ql_{\rm A})}\right)
\times \\ \left(\frac{1}{M_{\rm B}}\frac{
1-(j_0(ql_{\rm B}))^{M_{\rm B}}}{1-j_0(ql_{\rm B})}\right)
\label{eq:FAB}
\end{multline} 
and
\begin{multline} 
F_{\rm B,B}=
\frac{1}{M_{\rm B}^2}\left(M_{\rm B}+ \right.\\ \left.
\frac{2}{1-j_0(ql_{\rm B})}
\left\{
(M_{\rm B}-1)-\frac{j_0(ql_{\rm B})-(j_0(ql_{\rm B}))^{M_{\rm B}}}{1-j_0(ql_{\rm B})}
\right\}\right).
\nonumber
\end{multline} 
Again there are two solutions for this $2\times 2$ eigenvalue problem but this time
the plus sign (see again Eq.~\ref{eq:eigenvaluenem}) yields the physical bifurcation
density, $\tilde{n}_{\rm mps}$, for the microseparated phase (mps),
\begin{multline}
\tilde{n}_{\rm mps}=-\frac{4(1+\tilde{M})(1+\tilde{M}\tilde{l}^2\tilde{d})}{\pi M}
\times \\
\left(\tr({\bf F}(q){\bf G}_0)+
\sqrt{\tr^2({\bf F}(q){\bf G}_0)
-4\det{\bf F}(q)\det{\bf G}_0}
\right)\\ /(2 \det{\bf F}(q)\det{\bf G}_0).
\label{eq:eigenvaluemps}
\end{multline}
Apart from the approximations made in formulating the model, Sec.~\ref{sec:model},
Eqs.~\ref{eq:approx1} and~\ref{eq:approx23} constitute the only two further approximations.
From Eq.~\ref{eq:eigenvaluemps} it is observed directly that the spinodal density of the microseparated
phase scales with $1/M$, contrary to the nematic spinodal, Eq.~\ref{eq:eigenvaluenem2} which does not
depend on $M$ in this representation.
Consequently, for long enough polymers the system will always become unstable with respect to the
microseparated phase.
Furthermore, we note that for infinitely long chains ($M\rightarrow\infty$) the approximations become
exact (and the density needs to be rescaled, $\tilde{n}M$).
If the chains are not long, the approximations, Eqs.~\ref{eq:approx1} and~\ref{eq:approx23} will not be valid.
An interesting case are e.g.\ rod-coil copolymers where $M_{\rm A}=1$ and $M_{\rm B}$ is large.
The type-A rods will tend to form a smectic which the type-B tails are likely to stabilize~\cite{duijneveldt97,casey99}.
In this case, Eq.~\ref{eq:bifurcation4} has to be solved numerically or in some other (approximate) way.
Moreover, the ordering of the type-A rods is then likely to be dominated by an orientationally ordered
density fluctuation, e.g.\ possibly $\exp[i{\bf q}\cdot{\bf r}]P_2(\hat{q}\cdot\ho)$, instead of
the simple $\exp[i{\bf q}\cdot{\bf r}]$ which we have in the present case.
Finally, we note that the specification of the geometry is contained in the matrix ${\bf F}(q)$.
Using other geometries, e.g.\ ABABAB... repeating multiblock copolymers or branched geometries,
do not change Eqs.~\ref{eq:bifurcation4} or~\ref{eq:eigenvaluemps} but only the form of ${\bf F}(q)$
(the only requirement is that there are no closed loops within the polymers~\cite{wessels04}).

\section{The Gaussian Limit}
\label{sec:hdgl}

In this section we will construct a consistent limit for
infinitely long chains of our model. There are several reasons for
this approach. First of all, there is a large body of literature
dealing with so-called Gaussian chains, i.e.\ polymers which are
coarse-grained on the level of the radius of gyration, and we want
to make contact with those treatments~\cite{bates90,leibler80}.
Secondly, we do not fully control the quality of the
approximations, Eqs.~\ref{eq:approx1} and~\ref{eq:approx23}, made
for chains of finite length. It is clear, however, that these
approximations become exact for infinitely long polymers. Finally,
by introducing this limiting case the number of effective model
parameters is reduced, resulting in a conceptually simpler system.
The limit of $M_{\rm A},M_{\rm B}\rightarrow\infty$ does require
that some of the other parameters be rescaled as well.
Additionally, we want to take this limit in such a way that the
nematic and microseparated bifurcation densities remain of the
same order of magnitude so that we can compare them. This extra
requirement is non-trivial as can be seen from
Eqs.~\ref{eq:eigenvaluenem2} and~\ref{eq:eigenvaluemps} because
$\tilde{n}_{\rm mps}$ scales with $1/M$ and thus vanishes for long
polymers. We can cure this divergence in a somewhat unconventional
way by letting the difference in thickness of the A and B segments
vanish, $\tilde{d}\rightarrow 1$. In this way, the incentive for
MPS is much reduced and $\tilde{n}_{\rm mps}$ ``pulled up'' to
nonzero densities comparable to $\tilde{n}_{\rm nem}$.
Summarizing, we take the limits \be \ba{cccc} M_{\rm
A}\rightarrow\infty, & l_{\rm A}\rightarrow
0&\mbox{and}&\tilde{d}\rightarrow 1 \ea\ee whilst $M_{\rm
A}l^2_{\rm A}$ and $M_{\rm A}(1-\tilde{d})^2$ remain finite.
Furthermore, we keep the ratios $\tilde{M}$ and $\tilde{l}$ fixed,
such that the type-B segments are subject to the same limit. Next,
in order for the Onsager approximation to still be valid, $d_{\rm
A}$ needs to remain much smaller than $l_{\rm A}$ and therefore
needs to go to zero even faster. This is corrected by letting the
number density of chains go to infinity in order to keep total
strength of the interaction, i.e.\ the total excluded volume
constant. So additionally we have \be\ba{ccc} d_{\rm A}\rightarrow
0&\mbox{and}&n\rightarrow \infty \ea \ee with $2nM_{\rm A}l^2_{\rm
A}d_{\rm A}$ and therefore also $\tilde{n}=2n(M_{\rm A}l^2_{\rm
A}d_{\rm A}+M_{\rm B}l^2_{\rm B}d_{\rm B})$ finite.

In the Gaussian limit, the relevant length scale is the radius of
gyration or equivalently, the mean-square end-to-end distance. The
mean-square end-to-end distance is defined as \be
x^2=\sum_{k,k'}<l_k\ho_k\cdot l_{k'}\ho_{k'}>, \ee where $<>$
denotes the average over a single chain~\cite{degennes79}. In a
freely-jointed chain there is no orientational correlation between
the segments so for our block copolymers, the mean-square
end-to-end distance is simply $x^2=M_{\rm A}l^2_{\rm A}+M_{\rm
B}l^2_{\rm B}$. This allows us to define the dimensionless
wavenumber as $\tilde{q}=qx$.

Our reduced model has three parameters, $\tilde{M}$, $\tilde{l}$
governing the composition and the relative size of the copolymeric
blocks and $\tilde{\Delta} \equiv M_{A}(1-\tilde{d})^2$ describing
the remaining difference in thickness between the two components
and hence effectively setting the incentive for demixing.

In the Gaussian limit, the determinant of ${\bf G}_2$ goes to
zero, $\det {\bf G}_2=
-\frac{1}{4}\tilde{M}\tilde{l}^2(1-\tilde{d})^2\rightarrow 0$.
Consequently, we can expand Eq~\ref{eq:eigenvaluenem2} for small
$\det {\bf G}_2$ and we obtain for the nematic bifurcation density
in the Gaussian limit, \be n_{\rm
nem}=\frac{32(1+\tilde{M}\tilde{l}^2)}{\pi\tr{\bf
G}_2(\tilde{d}=1)}= \frac{32}{\pi}, \label{eq:bifnemgauss} \ee
which, conveniently, is a constant independent on the model
parameters. Setting the first element of the eigenvector to one,
${\bf c}_{\rm nem}=(1,c_{\rm nem})$, this is very simple in the
Gaussian limit, $c_{\rm nem}=\tilde{l}$. Therefore, at the nematic
bifurcation the B segments are $\tilde{l}$ times more strongly
orientationally ordered than the A segments.

Concerning MPS, we first calculate the elements of ${\bf F}$ in
the Gaussian limit, \be F_{\rm A,A}=\frac{12}{q_{\rm
A}^2} \left\{ 1-\frac{6}{q_{\rm A}^2} \left(1-{\rm e}^{-q_{\rm
A}^2/6}\right) \right\} \nonumber \ee \be F_{\rm A,B}=F_{\rm B,A}=
\frac{6}{q^2_{\rm A}}\left(1-{\rm e}^{-q_{\rm A}^2/6}\right)
\frac{6}{q^2_{\rm B}}\left(1-{\rm e}^{-q_{\rm B}^2/6}\right)
\label{eq:Felemgauss} \ee \be F_{\rm B,B}=\frac{12}{q_{\rm B}^2}
\left\{ 1-\frac{6}{q_{\rm B}^2} \left(1-{\rm e}^{-q_{\rm
B}^2/6}\right) \right\} \nonumber \ee with \be\ba{ccc} q_{\rm
A}^2={\displaystyle
\frac{\tilde{q}^2}{1+\tilde{M}\tilde{l}^2}}&\mbox{and}&q_{\rm
B}^2= {\displaystyle
\frac{\tilde{q}^2\tilde{M}\tilde{l}^2}{1+\tilde{M}\tilde{l}^2}}
\ea\ee The determinant of ${\bf G}_0$ also goes to zero, $\det
{\bf G}_0=
-\frac{1}{4}\tilde{M}^2\tilde{l}^2(1-\tilde{d})^2\rightarrow 0$.
Next, expanding Eq.~\ref{eq:eigenvaluemps} for small $\det {\bf
G}_0$ as well, we obtain for the bifurcation density of MPS in the
Gaussian limit, 
\begin{multline} \tilde{n}_{\rm
mps}=-\lim_{\tilde{d}\rightarrow
1}\frac{4(1+\tilde{M})(1+\tilde{M}\tilde{l}^2\tilde{d})}{\pi M}
\frac{\tr({\bf F}{\bf G}_0)}{\det{\bf F}\det{\bf G}_0} \\ =
\frac{16(1+\tilde{M})(1+\tilde{M}\tilde{l}^2)}{\pi
\tilde{\Delta}^2\tilde{M}^2\tilde{l}^2} \frac{\tr({\bf
F}\tilde{\bf G}_0)}{\det{\bf F}}, \label{eq:bifmspgauss} \end{multline} 
with
$\tilde{\bf G}_0=\lim_{\tilde{d}\rightarrow 1}{\bf G}_0$, \be
\tilde{\bf G}_0=\left[\ba{cc}1 & \tilde{M}\tilde{l}\\
\tilde{M}\tilde{l} & \tilde{M}^2\tilde{l}^2 \ea\right].
\label{eq:tildeG0} \ee
Additionally, we note the symmetry in the A
and B types, i.e.\ the following transformation
$\{\tilde{\Delta},\tilde{M},\tilde{l}\}\rightarrow\{\tilde{\Delta},1/\tilde{M},1/\tilde{l}\}$
leaves the results unchanged. Again, writing the eigenvector as
follows ${\bf c}_{\rm mps}=(1,c_{\rm mps})$ we obtain a simple
expression in the Gaussian limit $c_{\rm
mps}=-1/\tilde{M}\tilde{l}$. This is the relative order 
of component B over A at bifurcation. The
minus sign is due to the difference of $\pi$ in phase between the
density waves of A and B, i.e. where the density of A is enhanced
the density of B is depressed (${\rm e}^{i\pi}=-1$). The absolute
value $1/\tilde{M}\tilde{l}$ is ratio of amplitudes of the two
waves. The matrix ${\bf F}$ contains the correlations within the
polymer and is seen to feature the so-called Debije functions,
$g_D(x)=(2/x)(1-(1/x)(1-\exp[-x]))$ reflecting the Gaussian
character of the correlations. In the Leibler
approach~\cite{leibler80} these appear in a similar way and
therefore, the correlations are treated on the same level.

\section{Results}
\label{sec:results}

\subsection{Bifurcation Density}
\label{sec:resbifdens}

\begin{figure}[t]
\epsfig{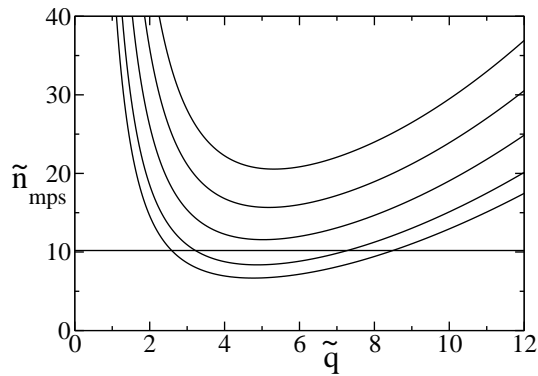}
\caption{\small Bifurcation density for the microseparated phase vs.\
the magnitude of the wave vector for $\tilde{l}=1$ and $\tilde{\Delta}=4$ and $\tilde{M}=\{5,4,3,2,1\}$
(from top to bottom).
The nematic bifurcation density $\tilde{n}_{\rm nem}=32/\pi\approx 10$ and has no wave dependence but is drawn as a
straight line for comparison.
Due to symmetry ($\{\tilde{M},\tilde{l}\}\rightarrow\{1/\tilde{M},1/\tilde{l}\}$) the curves of $\tilde{M}$
are the same for $1/\tilde{M}$.
}
\label{fig:etaq}
\end{figure}
In Fig.~\ref{fig:etaq}, we have plotted the analytical bifurcation
density of the microseparated phase, Eq.~(\ref{eq:bifmspgauss}) as
a function of the wave vector for various values of $\tilde{M}$.
Most importantly, all curves have a minimum for a certain wave
vector. Interpreting the bifurcation point as the spinodal, where
the isotropic fluid phase changes from being stable to unstable,
the system becomes first unstable for fluctuations with a wave
length corresponding to the minimum density. We have also plotted
the nematic bifurcation density, being a constant independent of
the wavenumber $\tilde{q}$, in Fig.~\ref{fig:etaq}. For the curves
which lie totally above the horizontal line, the system becomes
unstable with respect to the nematic phase at the density
$\tilde{n}=\tilde{n}_{\rm nem}=32/\pi$. For a curve of which
the minimum reaches below the horizontal line, the system becomes
unstable with respect to a microseparated phase with wave length
$\tilde{\lambda}_{\rm min}=2 \pi /\tilde{q}_{\rm min}$ at the
minimum density $\tilde{n}=\tilde{n}_{\rm mps}^{({\rm
min})}$. In Fig.~\ref{fig:etaq}, we have set the A and B segments
to equal length, $\tilde{l}=1$ and the demixing parameter is
$\tilde{\Delta}=4$ . Starting with an asymmetric polymer,
$\tilde{M}=5$, MPS only occurs for high densities. Making the
polymer more symmetric and decreasing $\tilde{M}$ to one, the
curves shift to lower densities until at $\tilde{M}=1$ it is at
its lowest position. Upon a further decrease $\tilde{M}$ following
the sequence
$\tilde{M}=\{1,\frac{1}{2},\frac{1}{3},\frac{1}{4},\frac{1}{5}\}$,
we again follow the same curves in Fig.~\ref{fig:etaq} due to the
symmetry
$\{\tilde{\Delta},\tilde{M},\tilde{l}\}\rightarrow\{\tilde{\Delta},1/\tilde{M},1/\tilde{l}\}$
and the choice $\tilde{l}=1$, but now from the bottom to the top.

\begin{figure}[t]
\epsfig{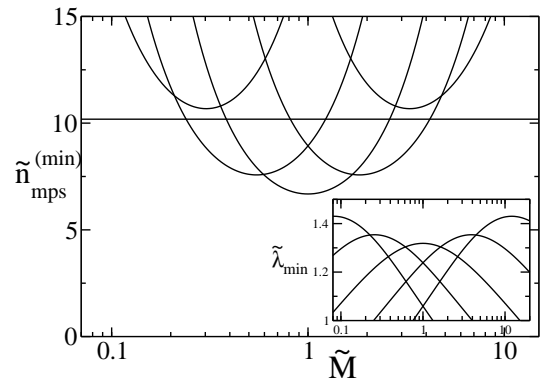}
\caption{\small The minimum bifurcation density for the microseparated phase vs.\
$\log\tilde{M}$ for $\tilde{\Delta}=4$ and $\tilde{l}=\{0.25,0.5,1,2,4\}$ (right to left).
The nematic bifurcation density $\tilde{n}_{\rm nem}=32/\pi\approx 10$ is constant
and drawn as a straight line for comparison.
Inset: the wave length for which the bifurcation density of the microseparated phase is a minimum,
$\tilde{\lambda}_{\rm min}=2\pi/\tilde{q}_{\rm min}$ vs.\ $\tilde{M}$ for the same parameters;
$\tilde{\Delta}=4$ and $\tilde{l}=\{0.25,0.5,1,2,4\}$ (right to left).
}
\label{fig:etaminM}
\end{figure}
We have numerically determined the minimum of the MPS bifurcation
density with respect to the wave vector,
Eq.~(\ref{eq:bifmspgauss}) and plotted that in
Fig.~\ref{fig:etaminM} as a function of $\tilde{M}$ for a few
different $\tilde{l}$. We observe the same trend we saw in
Fig.~\ref{fig:etaq}: for very asymmetric polymers, $\tilde{M}\ll
1$, the minimum MPS bifurcation density is very high. Increasing
$\tilde{M}$, the bifurcation density goes down until a certain
value $\tilde{M}$ (depending on $\tilde{l}$) after which it goes
up again. As shown in Fig.~\ref{fig:etaminM} some of the curves
reach below the horizontal line marking the stability limit of the
isotropic phase towards nematic ordering. Consequently, in the
intermediate region the microseparated phase is probably the most
stable phase, whereas for the more asymmetric polymers MPS
is likely to be preempted by the nematic
phase. Furthermore, there is also a dependence on $\tilde{l}$,
i.e.\ increasing the asymmetry between the A and B segments, the
curves shift to higher densities. Again, we note that the two
curves for $\tilde{l}=0.5$ and $\tilde{l}=2$ can be mapped onto
each other due to symmetry in the model parameters. In the inset
of Fig.~\ref{fig:etaminM} we have plotted the value of the wave
length $\tilde{\lambda}=2\pi/\tilde{q}$ corresponding to
$\tilde{n}_{\rm mps}^{({\rm min})}$ vs.\ $\tilde{M}$. There is
a rough correspondence as a function of $\tilde{M}$ in that the
lower the MPS bifurcation densities in Fig.~\ref{fig:etaminM}
connect to the higher wave lengths in
Fig.~\ref{fig:etaminM}(inset). In general, we have observed that
the wave length for which the MPS is the stable phase (over the
nematic) roughly lie between 1 and 1.5 times the mean end-to-end
distance $x$, i.e.\ the polymers get somewhat stretched at the
phase transition.

\subsection{Phase Diagrams}

\begin{figure}[t]
\epsfig{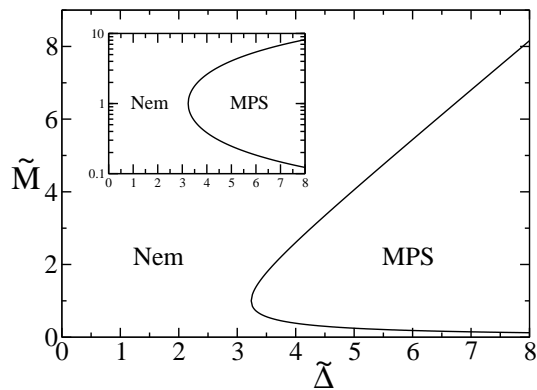}
\caption{\small Phase diagram, $\tilde{M}$ vs.\ $\tilde{\Delta}$ for
$\tilde{l}=1$.
For the region marked with ``Nem'', the lowest bifurcation density is the nematic
and for the region marked with ``MPS'' this is the microseparated phase.
The inset is the same phase diagram except that the vertical axis is logarithmic
to show the symmetry with respect to $\{\tilde{M},\tilde{l}\}\rightarrow\{1/\tilde{M},1/\tilde{l}\}$.
}
\label{fig:phaseMDel}
\end{figure}

\begin{figure}[t]
\epsfig{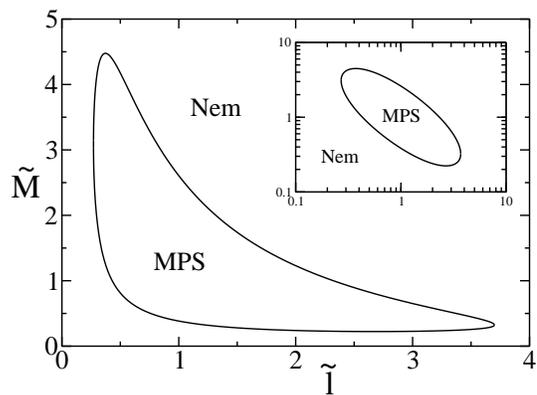}
\caption{\small  Phase diagram, $\tilde{M}$ vs.\ $\tilde{l}$ for
$\tilde{\Delta}=4$.
For the region marked with ``Nem'', the lowest bifurcation density is the nematic
and for the region marked with ``MPS'' this is the microseparated phase.
The inset is the same phase diagram except that the axes are logarithmic
to show the symmetry with respect to $\{\tilde{M},\tilde{l}\}\rightarrow\{1/\tilde{M},1/\tilde{l}\}$.
}
\label{fig:phaseMlDel4stepp01}
\end{figure}

Figs.~\ref{fig:phaseMDel} and~\ref{fig:phaseMlDel4stepp01} present
the phase diagrams. We have numerically computed the model
parameters for which the minimum MPS bifurcation density equals
the nematic bifurcation density. In Fig.~\ref{fig:phaseMDel}, the
phase diagram is given in terms of $\tilde{M}$ vs.\
$\tilde{\Delta}$ for equal length segments, $\tilde{l}=1$. For low
$\tilde{\Delta}$ the incentive for MPS is too weak and the MPS
bifurcation densities are higher than the nematic ones everywhere.
Increasing $\tilde{\Delta}$, the MPS becomes stable for
$\tilde{M}=1$ (totally symmetric diblock copolymer) and increasing
$\tilde{\Delta}$ further the range of $\tilde{M}$ for stable MPS
grows correspondingly. This is not surprising as the MPS
bifurcation density scales simply with $1/\tilde{\Delta}$. The
inset of Fig.~\ref{fig:phaseMDel} shows the vertical scale
logarithmically to show the symmetry with respect to
$\tilde{M}\rightarrow 1/\tilde{M}$. In
Fig.~\ref{fig:phaseMlDel4stepp01}, the phase diagram is plotted
for $\tilde{M}$ vs.\ $\tilde{l}$. The same observation as in
Sec.~\ref{sec:resbifdens} can be made: for asymmetric polymers,
the nematic phase is the most stable whereas for more symmetric
ones the MPS can be stable. Of course the amount of area in
Fig.~\ref{fig:phaseMlDel4stepp01} depends sensitively on
$\tilde{\Delta}$. Note that $\tilde{l}$ plays a very similar role
as $\tilde{M}$. Naively, one might expect that a difference in
lengths of the segments would also increase the tendency to
microphase separate or at least not counteract to it. However,
this is not the case, and only the difference in thickness, even
though only infinitesimally small in the Gaussian limit, drives
the occurence of MPS, in line with earlier work on binary mixtures
of rods~\cite{vanroij96}. Potentially, length differences between
the component rods could drive MPS within the nematic phase, but
probing this would require the numerical solutions to the full
self-consistency problem, currently beyond our scope.

\subsection{The Density Shift along the Polymer}

\begin{figure}[t]
\epsfig{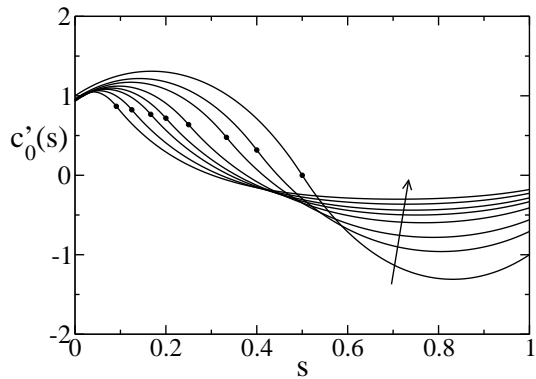}
\caption{\small Relative order along the polymer at
bifurcation in the microseparated phase, $c'^{(0)}_{\rm A}(s)$ for
$s\in [0,1/(1+\tilde{M})]$ and $c'^{(0)}_{\rm B}(s)$ for $s\in
[1/(1+\tilde{M}),1]$. Parameters are $\tilde{\Delta}=4$,
$\tilde{l}=1$ and $\tilde{M}=\{1,1.5,2,3,4,5,7,10\}$ (increasing
in the direction of the arrow). The normalization is such that the
averages over $c'^{(0)}_{\rm A}(s\in {\rm A})$ and $c'^{(0)}_{\rm
B}(s\in {\rm B})$ equal $c^{(0)}_{\rm A}=1$ and  $c^{(0)}_{\rm
B}=-1/(\tilde{M}\tilde{l})$ respectively. The full circles
indicate the ``joints'' of the A and B parts at
$s=1/(1+\tilde{M})$. } \label{fig:OrderalongM}
\end{figure}

\begin{figure}[t]
\epsfig{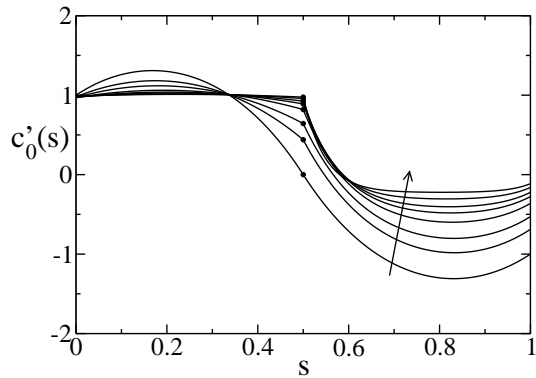}
\caption{\small Relative order along the polymer at
bifurcation in the microseparated phase, $c'^{(0)}_{\rm A}(s)$ for
$s\in [0,1/(1+\tilde{M})]$ and $c'^{(0)}_{\rm B}(s)$ for $s\in
[1/(1+\tilde{M}),1]$. Parameters are $\tilde{\Delta}=4$,
$\tilde{M}=1$ and $\tilde{l}=\{1,1.5,2,3,4,5,7,10\}$ (increasing
in the direction of the arrow). The normalization is such that the
averages over $c'^{(0)}_{\rm A}(s\in {\rm A})$ and $c'^{(0)}_{\rm
B}(s\in {\rm B})$ equal $c^{(0)}_{\rm A}=1$ and  $c^{(0)}_{\rm
B}=-1/(\tilde{M}\tilde{l})$ respectively. The full circles
indicate the ``joints'' of the A and B parts at
$s=1/(1+\tilde{M})$. } \label{fig:Orderalongl}
\end{figure}

The elements of the eigenvectors at the bifurcation as discussed
in Secs.~\ref{sub:nematicsolution} and~\ref{sub:mspsolution}
contain information about the relative amplitude of the nascent
ordering with respect to the the homogeneous and isotropic parent
phase. However, by construction these quantities were averaged
over all segments either of type A or B. In case of the nematic
ordering, this also coincides exactly with the order of each of
the segments individually as there is no orientational coupling
between the segments and these therefore behave as being
independent. However, in case of MPS, there clearly is a spatial
coupling between the segments and, consequently, one would expect
a different degree of ordering e.g.\ for segments which are close
to the free end and those which are close to the joint. Those
close the joint are be subjected to two counteracting density
waves and will order less than those at the free ends. In order to
to quantify these effects we have to compute the components of the
$M$-dimensional vector $c_m^{(0)}$ (Eq.~(\ref{eq:bifurcation6})).
In appendix~\ref{sec:orderwithin}, we explain how these are
obtained from the type-averaged 2-dimensional eigenvectors by
means of an additional quantity: the half type-averaged matrix
${\bf F}'$. In the Gaussian limit, this $M$-dimensional vector
reduces to the following  2-dimensional eigenvector (with a prime), \be
{\bf c}'_{0}(s)=\left(\ba{c}c'^{(0)}_{\rm A}(s\in A)\\
c'^{(0)}_{\rm B}(s\in B)\ea\right). \ee which now depends, on the
continuous label $s\in[0,1]$, where
$s\in[0,\frac{1}{1+\tilde{M}}]$ implies $s\in {\rm A}$ and
$s\in[\frac{1}{1+\tilde{M}},1]$ implies $s\in {\rm B}$. In
Figs.~\ref{fig:OrderalongM} and~\ref{fig:Orderalongl}, we plot the
components of ${\bf c}'_{0}(s)$ along the polymer (as a function
of $s$) for increasing $\tilde{M}$ and $\tilde{l}$ respectively.
The demixing parameter is taken to be $\tilde{\Delta}=4$.

In Fig.~\ref{fig:OrderalongM}, we start from the symmetric case,
$\tilde{M}=1$ and $\tilde{l}=1$ where the profile is also
symmetric around $s=0.5$. All A segments have positive order and
all B segments have negative order and the average of A and B is
+1 and $-1$ respectively as expected. Increasing  $\tilde{M}$, the
B part of the polymer becomes larger than the A part and the joint
shifts to the left. The normalization remains such that average
order of the A segments is still 1 and that of the B segments is
$-1/\tilde{M}$. However, it is remarkable that the B segments
close to the joint obtain a positive order with increasing
$\tilde{M}$, i.e.\ they order with respect to the density wave of
A instead of that of B. This is due to the fact that in the
polymer there is much more material from the B part. Consequently,
this effect becomes stronger for larger $\tilde{M}$. In
Fig.~\ref{fig:Orderalongl}, we start again from the symmetric
case, $\tilde{M}=1$ and $\tilde{l}=1$. Subsequently, the ratio of
lengths $\tilde{l}$ is increased and we see that the derivative of
the profile to $s$ jumps at the joint. Furthermore, also here, the
joint shifts to positive values and the A segments have a much
more constant profile than the B segments. By increasing
$\tilde{l}$ while $\tilde{M}$ remains constant one effectively
increases the amount of material in the B part of the polymer.
Therefore, it is not surprising that the point of zero order
shifts to the right. Additionally, the B segments are much longer
and therefore the spatial correlations persist over larger $s$
explaining the more smooth profile on the B side. It has to be
noted that some of the profiles (especially for higher values of
$\tilde{M}$ and $\tilde{l}$ in Figs.~\ref{fig:OrderalongM}
and~\ref{fig:Orderalongl}) are taken at bifurcation densities far
above the nematic bifurcation. We have nevertheless included them,
being instructive in explaining the observed trends.

\section{Conclusion}
\label{sec:conclusion}

We have considered a fluid of freely-jointed hard diblock
copolymers. The two polymer blocks A and B consist of slender
Onsager rods of different dimensions interacting via hard body
repulsion only. We apply a DFT approach in the second virial
approximation from first principles, and analytically construct
local solutions to the stationarity equations, by means of a
stability (bifurcation) analysis of the isotropic phase. Spatial
as well as orientational degrees of freedom are taken into account
and consequently we obtain the spinodal densities for both the
microseparated and the nematic phases. It is shown that for long
polymers the system always becomes unstable with respect to the
microseparated phase first. Consequently, this means that entropy
can induce MPS in much the same way as it has been found to induce
other forms of spontaneous ordering before. Furthermore, the
mechanism is determined solely by the (difference in) dimensions
of the rods and therefore has a conceptually simple geometric
origin.

In order to make contact with the literature on thermotropic block
copolymers we take the limit of infinitely long polymers in which
the approximations become exact. In addition, by assuming a
vanishing difference in thickness of the two types of rods, we can
still study the competition of the microseparated with the nematic
phase. We present phase diagrams in terms of model parameters
showing the regions of stable microseparated or nematic ordering.
We also present the order along the polymer at the bifurcation of
the microseparated phase.

In the present study, we have solved the stationarity equations up
to first-order in a bifurcation analysis. This yields, apart from
the location of the spinodal or bifurcation density, only the
magnitude of the density wave vector and the spherical harmonic 
mode to which the isotropic solution becomes
unstable. However, the symmetry of the bifurcating microseparated
solution is typically determined by one or more mutually
independent (but equally long) vectors spanning the periodic phase
(e.g.\ lamellar, hexagonal or bcc). In order to obtain information
on the mutual orientation of these lattice vectors, and thus on
the symmetry of the phase, a higher order bifurcation analysis
should be performed~\cite{kayser78,mulder87,mulder89}. From these
higher order bifurcation equations, it is also possible to
determine whether the phase transition is of first or second order
and in the latter case one could in principle go on to construct
the full equilibrium solution far away from the bifurcation
point~\cite{kayser78}.

We have not checked the validity of the approximations,
Eqs.~\ref{eq:approx1} and~\ref{eq:approx23} for finite values of
$M$. However, we can make a crude estimate, a posteriori, by
concluding from Fig.~\ref{fig:etaminM} that the bifurcating wave
length is of the order of the mean square end-to-end distance,
$\tilde{\lambda}=\lambda/x\sim 1$. Consequently, the wave vector
is approximately, $\tilde{q}=2\pi/\tilde{\lambda}\sim 2\pi$ and if
we assume for a moment that the type-A rods and type-B rods are
more or less equally long, then the mean-square end-to-end
distance is $x^2\sim M l_{\rm A}^2$. This in turn implies that the
next order corrections in Eqs.~\ref{eq:approx1}
and~\ref{eq:approx23} will be of order $(\frac{1}{2}ql_{\rm A})^2
\sim (\frac{1}{2} \times 2\pi)^2/M\sim 10/M$. (In fact the first
order correction in $(\frac{1}{2}ql_{\rm A})$ in
Eq.~\ref{eq:approx23} does not contribute to the value of the
bifurcation density, but only to the form of the eigenfunction.)
Consequently, already for this crude test case, the length of the
polymer should be \emph{at least} longer than 10 ($M>10$) in order
for the corrections to be smaller than the leading term. This
suggests that much higher values of $M$ are required for the
present approach to yield quantitative agreement with the "true"
behaviour.

In any case, it would be very interesting to extend the present
approach to finite values of $M$. However, this is not
straightforward, as the correlations within the chain would become
non-Gaussian. One strategy could be to solve
Eq.~\ref{eq:bifurcation4} directly numerically but this could
become tedious for large numbers of segments. Another strategy
would be to make an expansion in $1/M$ using the Gaussian limit as
a reference state. This last route was followed by Fredrickson and
Helfand~\cite{fredrickson87} for Leibler's diblock copolymers and
the results were confirmed by simulations~\cite{fried91}. Indeed,
there is a need for such a better-than-Gaussian treatment,
especially when the typical ordering length scales are of the same
sizes as the components, e.g.\ for side chain liquid crystalline
polymers forming a smectic~\cite{donald92,renz86}.

As already mentioned in the introduction, there is as yet no
experimental system exhibiting MPS due to the mechanism described
in this paper. However, considering the ongoing progress in the
field of bio-engineering~\cite{dogic01,surrey01}, it may become
possible to prepare such a system. We mention again the
possibility of long and thin polymers connected to TMV rods in an
appropriate solvent. The solvent may be a problem as we have the
double requirement that the polymers are at their $\theta$-point
and that at the same time the TMV rods still act as hard
particles. Still, such a system of entropic rod-coil copolymers
could be  directly compared to the simulation studies of
Refs.~\cite{duijneveldt97,casey99}. Additionally, it would be
described by Eq.~\ref{eq:bifurcation4}, which would than have to
be solved for the case of $M_{\rm A}=1$ and $M_{\rm B}$ large. In
a more general context,it becomes increasingly clear that
entropy-induced effects play a prominent role in
vivo~\cite{rizvi03}, and it may be that similar mechanisms as
described here prevent demixing tendencies due to local
constraints~\cite{surrey01}. On the other hand, the mechanism may
also be of relevance in thermotropic systems where the two
components of block copolymers also have short-range anisotropic
repulsions which are usually of different range. In any case,
observing entropy-induced microphase separation in monodisperse
systems would certainly be an interesting experimental challenge.


\begin{acknowledgments}
The authors would like to thank M.\ Schmidt and D.\ Lukatsky for critically reading the manuscript. 
This work is part of the research program of the ``Stichting voor Fundamenteel Onderzoek der Materie'' (FOM),
which is financially supported by the ``Nederlandse Organisatie voor Wetenschappelijk Onderzoek'' (NWO).
PPFW would like to thank the Heinriche-Heine-Universit\"{a}t D\"{u}sseldorf for hospitality, where part of writing this paper
was finished.
\end{acknowledgments}

\bibliographystyle{unsrt}
\bibliography{paulslit}

\appendix
\section{The Fourier Transformed Segment-Segment Mayer Function}
\label{sec:ftmayer}

The Mayer function $\phi_{k,k'}$ of two cylindrical rodlike segments $k$ (with dimensions
$l_k,d_k$ and coordinates $({\bf r}_k,\ho_k)$) and $k'$ (with $l_{k'},d_{k'}$ and
$({\bf r}_{k'},\ho_{k'})$) interacting
via a hard-core potential (i.e.\ $=\infty,0$ if overlap/no overlap)
is given by
\be
\phi_{k,k'}({\bf r}_k-{\bf r}_{k'},\ho_k,\ho_{k'})=
\left\{ \ba{lll}-1&\mbox{if overlap}\\
0&\mbox{if no overlap}\ea\right.
\ee
We decompose the spatial vector ${\bf r}_{k,k'}={\bf r}_k-{\bf r}_{k'}$
in terms of the orientations,
\be
{\bf r}_{k,k'}=x_k\ho_k+x_{k'}\ho_{k'}+x_{k,k'}\ho_{k,k'}
\ee
with $\ho_{k,k'}=(\ho_k\times\ho_{k'})/|\ho_k\times\ho_{k'}|$
the unit vector in the perpendicular direction.
There is overlap between the two rods for the following ranges of the coefficients,
$x_k\in [-l_k/2,l_k/2]$, $x_{k'}\in [-l_{k'}/2,l_{k'}/2]$ and
$x_{k,k'}\in [-(d_k+d_{k'})/2,(d_k+d_{k'})/2]$.
Next, the Fourier transform of the Mayer function $\hat{\phi}_{k,k'}$ is given by
\be
\hat{\phi}_{k,k'}({\bf q},\ho_k,\ho_{k'})=
\int d{\bf r}_{k,k'} {\rm e}^{-i {\bf q}\cdot {\bf r}_{k,k'}}
\phi_{k,k'}({\bf r}_{k,k'},\ho_k,\ho_{k'}),
\ee
where the volume of the infinitesimal element is given by
$d{\bf r}_{k,k'}=|\ho_k\times\ho_{k'}|dx_k dx_{k'} dx_{k,k'}$.
Consequently,
\begin{multline}
\hat{\phi}_{k,k'}({\bf q},\ho_k,\ho_{k'})= \\
-|\ho_k\times\ho_{k'}| \int_{-l_k/2}^{l_k/2}dx_k
\int_{-l_{k'}/2}^{l_{k'}/2}dx_{k'}
\int_{-(d_k+d_{k'})/2}^{(d_k+d_{k'})/2}dx_{k,k'} \times \\
\exp[-i(x_k{\bf q}\cdot\ho_k+x_{k'}{\bf q}\cdot\ho_{k'}+x_{k,k'}{\bf q}\cdot\ho_{k,k'})]
\\
=-l_{k}l_{k'}(d_{k}+d_{k'})|\ho_{k}\times\ho'_{k'}|
j_0\left({\textstyle \frac{1}{2}}l_k {\bf q}\cdot\ho_k\right)\times \\
j_0\left({\textstyle \frac{1}{2}}l_{k'} {\bf q}\cdot\ho'_{k'}\right)
j_0\left({\textstyle \frac{1}{2}}(d_k+d_{k'}) {\bf q}\cdot\ho_{k,k'}\right)
\end{multline}
with the spherical Bessel function of zeroth order given by
$j_0(x)=\sin x/x$.
In the Onsager limit of very slender rods, $l_k,l_{k'}\gg d_k,d_{k'}$
while $l_{k}l_{k'}(d_{k}+d_{k'})$ stays finite.
In our system, we expect the wave length of the microseparated phase
to be {\em at least} of the order of the lengths of the
segments (although for large number of segments it is even much larger).
Consequently, in this case, $|(d_k+d_{k'}){\bf q}|\ll 1$ and
we use the leading order, which is
$j_0({\textstyle \frac{1}{2}}(d_k+d_{k'}) {\bf q}\cdot\ho_{k,k'})=1$.
Then, our final result for the Mayer function is
\begin{multline}
\hat{\phi}_{k,k'}({\bf q},\ho_k,\ho_{k'})
=-l_{k}l_{k'}(d_{k}+d_{k'})|\ho_{k}\times\ho'_{k'}| \times \\
j_0\left({\textstyle \frac{1}{2}}l_k {\bf q}\cdot\ho_k\right)
j_0\left({\textstyle \frac{1}{2}}l_{k'} {\bf q}\cdot\ho'_{k'}\right).
\end{multline}

\section{The Eigenfunctions of $\hat{\phi}_{k,k'}(\ho\cdot\ho')$ for ${\bf q}=0$}
\label{sec:eigenfunctions}

For ${\bf q}=0$, the Fourier transformed Mayer function is
\begin{align}
\hat{\phi}_{k,k'}(\ho\cdot\ho')&=-l_kl_{k'}(d_k+d_{k'})|\ho\times\ho'| \\
&=-l_kl_{k'}(d_k+d_{k'})\sqrt{1-(\ho\cdot\ho')^2}
\end{align}
and is therefore uniaxial, i.e.\ dependent on a single planar angle
$\gamma=\arccos(\ho\cdot\ho')$.
Therefore, we can expand it in terms of Legendre polynomials
\be
\hat{\phi}_{k,k'}(\ho\cdot\ho')=-l_kl_{k'}(d_k+d_{k'})
\sum_{j=0}^{\infty}\frac{2j+1}{4\pi}s_jP_j(\ho\cdot\ho'),
\ee
with $s_j=2\pi\int_{-1}^{1} dx \sqrt{1-x^2}P_j(x)$.
Then, using the decomposition in terms of spherical harmonics $Y_{j,i}$,
we can rewrite this as
\begin{multline}
\hat{\phi}_{k,k'}(\ho\cdot\ho')=-l_kl_{k'}(d_k+d_{k'}) \times \\
\sum_{j=0}^{\infty}\sum_{i=-j}^{j}\frac{2j+1}{4\pi}s_jY_{j,i}(\ho\cdot\hat{z})Y_{j,i}^*(\ho'\cdot\hat{z}),
\end{multline}
with the asterisk denoting the complex conjugate and $\hat{z}$ some unit vector.
It is now directly seen that the Legendre polynomials are eigenfunctions of
$\hat{\phi}_{k,k'}(\ho\cdot\ho')$
\be
\int d\ho'\hat{\phi}_{k,k'}(\ho\cdot\ho')P_j(\ho'\cdot\hat{z})=
-l_kl_{k'}(d_k+d_{k'}) s_jP_j(\ho\cdot\hat{z}).
\ee

%

\section{Spatial Order within the Polymer}
\label{sec:orderwithin}

It is possible to calculate the bifurcating order within the polymer.
In case of freely-jointed chains in the nematic phase this is trivial
as this exactly $c_{\tau}^{(2)}$ for a segment of type $\tau$.
However, in case of MPS, segments of type A close to the ``joint''
with B segments will typically be more affected by the B part of the polymer than
segments of type A far away from the joint.
This order within the polymer can be obtained by calculating the elements of the $M$-dimensional vector
${\bf c}'_0$ with elements $c_m^{(0)}$ and $m\in\{1,\cdots ,M\}$ (see Eq.~\ref{eq:bifurcation6}).
Therefore we proceed by defining the matrix ${\bf F}'$ (with a prime)
\be
F'_{m\in\tau,\tau'}=\frac{1}{M_{\tau'}}\sum_{k'\in\tau'}F_{m,k'},
\ee
where the average is only performed over the second label
and therefore $F'_{m,\tau'}$ is $M\times 2$ dimensional.
Then, if the bifurcation density for the microseparated phase $\tilde{n}_{\rm mps}$ and the corresponding eigenvector
${\bf c}_{\rm mps}$ has been calculated beforehand (from Eq.~\ref{eq:eigenvaluemps}),
${\bf c}'_0$ can be computed by evaluating
\be
{\bf c}'_0=-{\displaystyle \frac{\pi\tilde{n}_{\rm mps}M}{4(1+\tilde{M})
(1+\tilde{M}\tilde{l}^2\tilde{d})}}{\bf F}'{\bf G}_0{\bf c}_{\rm mps}.
\ee
The elements of ${\bf F}'$ are given by
\begin{multline}
F'_{m\in{\rm A,A}}= \\ \frac{1}{M_{\rm A}}
\left(1+\frac{2-\left(j_0(ql_{\rm A})\right)^{m-1}-\left(j_0(ql_{\rm A})\right)^{M_{\rm A}-m}}{1-j_0(ql_{\rm A})}
\right)
\label{eq:FpAA}
\end{multline}
\begin{multline}
F'_{m\in{\rm A,B}}=  \frac{1}{M_{\rm B}}\left(j_0(ql_{\rm A})\right)^{M_{\rm A}-m}
\frac{1-\left(j_0(ql_{\rm B})\right)^{M_{\rm B}}}{1-j_0(ql_{\rm B})}
\label{eq:FpAB}
\end{multline}
\begin{multline}
F'_{m\in{\rm B,A}}= \frac{1}{M_{\rm A}}\left(j_0(ql_{\rm B})\right)^{m-M_{\rm A}-1}
\frac{1-\left(j_0(ql_{\rm A})\right)^{M_{\rm A}}}{1-j_0(ql_{\rm A})}
\label{eq:FpBA}
\end{multline}
\begin{multline}
F'_{m\in{\rm B,B}}= \\ \frac{1}{M_{\rm B}}
\left(1+\frac{2-\left(j_0(ql_{\rm B})\right)^{m-1-M_{\rm A}}-\left(j_0(ql_{\rm B})\right)^{M-m}}{1-j_0(ql_{\rm B})}
\right)
\label{eq:FpBB}
\end{multline}
where $m\in\{1,\cdots ,M_{\rm A}\}$ when $m\in A$ and $m\in\{M_{\rm A}+1,\cdots ,M\}$ when $m\in B$.
For each of these elements again holds that the average of $m$ yields the matrix ${\bf F}$ (see above
Eq.~\ref{eq:bifurcation7} and Eq.~\ref{eq:FAB}),
i.e.\
\be
F_{\tau,\tau'}=\frac{1}{M_{\tau}}\sum_{m\in\tau}F'_{m\in\tau,\tau'}.
\ee
In the Gaussian limit, we have to define a continuous ``label'', $s=m/M$, with $m$ and $M$ going to infinity
such that $s$ keeps its value.
Consequently,  $s\in[0,1]$ and ${\bf F}'$ becomes
\begin{multline}
F'_{\rm A,A}(s\in{\rm A})=\frac{6}{q^2_{\rm A}}
\left(2
-\exp\left[-\frac{q^2_{\rm A}}{6}s(1+\tilde{M})\right]
\right.\\ \left.
 -\exp\left[-\frac{q^2_{\rm A}}{6}(1-s(1+\tilde{M}))\right]\right)
\label{eq:FpAAgauss}
\end{multline}
\begin{multline}
F'_{\rm A,B}(s\in{\rm A})=\frac{6}{q^2_{\rm B}}
\left(1-\exp\left[-\frac{q^2_{\rm B}}{6}\right]\right) \times \\
\exp\left[-\frac{q^2_{\rm A}}{6}(1-s(1+\tilde{M}))\right]
\label{eq:FpABgauss}
\end{multline}
\begin{multline}
F'_{\rm B,A}(s\in{\rm B})=\frac{6}{q^2_{\rm A}}
\left(1-\exp\left[-\frac{q^2_{\rm A}}{6}\right]\right) \times \\
\exp\left[-\frac{q^2_{\rm B}}{6}(s\frac{1+\tilde{M}}{\tilde{M}}-\frac{1}{\tilde{M}})\right]
\label{eq:FpBAgauss}
\end{multline}
\begin{multline}
F'_{\rm B,B}(s\in{\rm B})=\frac{6}{q^2_{\rm B}}
\left(2
-\exp\left[-\frac{q^2_{\rm B}}{6}\left(s\frac{1+\tilde{M}}{\tilde{M}}-\frac{1}{\tilde{M}}\right)\right]
\right.\\ \left.
-\exp\left[-\frac{q^2_{\rm B}}{6}(1-s)\frac{1+\tilde{M}}{\tilde{M}}\right]\right)
\label{eq:FpBBgauss}
\end{multline}
where $s\in[0,\frac{1}{1+\tilde{M}}]$ when $s\in {\rm A}$
and  $s\in[\frac{1}{1+\tilde{M}},1]$ when $s\in {\rm B}$.
Note that in the Gaussian limit ${\bf F}'$ is simply a $2\times 2$ matrix, however, with $s$-dependence.
Consequently, unlike ${\bf F}$, ${\bf F}'$ is not symmetric.
And additionally, also the $M$-dimensional eigenvector becomes 2-dimensional,
\be
{\bf c}'_{0}(s)=\left(\ba{c}c'^{(0)}_{\rm A}(s\in A)\\ c'^{(0)}_{\rm B}(s\in B)\ea\right).
\ee
Finally, it has to be noted that in the Gaussian limit, first the product of ${\bf G}_0$ and ${\bf c}_{\rm mps}$ has to be taken and
only then the limit can be applied to $({\bf G}_0{\bf c}_{\rm mps})$.

\end{document}